# Estudo Abordando o Contexto de Notícias Falsas em Países de Língua Portuguesa (Fake News)

Carolina Duarte [0000-0002-7207-8769] e Valderi Leithardt [0000-0003-0446-9271]

VALORIZA, Research Center for Endogenous Resources Valoration, Instituto Politécnico de Portalegre, 7300-555 Portalegre, Portugal
email: {19206, valderi}@ipportalegre.pt

**Resumo.** Este trabalho consiste num estudo que aborda o contexto das fake news na realidade do mundo atual. Fake news é uma expressão muito utilizada atualmente. Durante o estudo realizado foi possível identificar problemas generalizados sobre esta temática, como por exemplo a grande propagação que estas têm e o impacto que apresentam na sociedade. A partir desses problemas foi possível identificar outros mais específicos, tais como a origem da notícia, a fonte da notícia, a pessoa que partilha e/ou cria a notícia e a relação interpessoal existente. Com a identificação deste subproblemas supracitados foi possível desenvolver um modelo taxonómico com o objetivo de implementar uma ferramenta que auxilie na deteção de fake news, identificando se uma notícia é verdadeira, falsa ou se o utilizador deve estar atento (quando não é possível identificar se a notícia é verdadeira ou falsa). Após a implementação, foi possível obter uma ferramenta que permite calcular a probabilidade de uma notícia ser falsa, considerando as opções selecionadas em cada parâmetro. Foi também possível verificar que a probabilidade estava corretamente calculada e que a ferramenta se revia no estudo realizado.

## 1 Introdução

Fake news é uma expressão muito utilizada atualmente, que ganhou maior visibilidade ao ser usada por Donald Trump quando se encontrava em campanha eleitoral para a presidência dos Estados Unidos da América em 2016. Sendo que por este motivo o termo foi considerado a palavra do ano de 2017 pelo dicionário Collins, referido em [1]. Apesar da sua grande utilização, não existe uma definição concreta para esta expressão, fazendo com que exista uma abundância de definições, o que pode gerar alguma preocupação relativa à heterogeneidade do termo "fake news", tal como indica [2].As fake news existem há milhares de anos tal como é possível comprovar através dos exemplos, conforme apresentados em [3]: No Império Romano, mais precisamente a 44 A.C., foi empreendida uma campanha contra Marco António por Otaviano com objetivo de denegrir a sua reputação; Em 1917, durante a Primeira Guerra Mundial foi realizada uma propaganda negativa, por parte do Reino Unido contra a Alemanha. Os jornais The Temple e The Daily Mail publicaram artigos indicando que devido à escassez de gordura na Alemanha, estes estariam a ferver os cadáveres dos soldados.

A evolução humana e tecnológica veio revolucionar a transmissão de informação, permitindo que as notícias falsas sejam disseminadas numa escala maior, tal como indica [4]. De acordo com o estudo "Influencia de las noticias falsas en la opinión pública", 88% dos entrevistados, acreditam que as notícias falsas são disseminadas de forma a prejudicar a imagem e reputação de pessoas ou organizações, sendo a também a reputação considerada o maior dano derivado das fake news.

Atualmente existe uma grande propagação de notícias falsas através de uma indústria especializada nessa ação. O exemplo dessa indústria é o caso da Media Vibes SNC, uma empresa belga que possui mais de 180 URLs dedicados à criação e divulgação de notícias falsas, como refere [6]. Os benefícios monetários associados a este tipo de indústria são uma das motivações para as pessoas se envolverem nestas atividades, comprovado pelo exemplo da referência [8], que indica que dezenas de adolescentes que publicaram notícias falsas, nas redes sociais durante as eleições presidenciais dos Estados Unidos da América em 2016, e enriqueceram devido aos cliques que obtiveram nessas notícias. Com a descoberta do Covid-19 e com a propagação do mesmo, iniciando uma pandemia mundial, originou-se também uma grande propagação de informações falsas, sendo estas partilhadas por milhares de pessoas que as consideram fidedignas, apesar de não o serem, conforme [7].

Associado a esta pandemia, está o termo infodemic, que é designado para delinear os perigos dos fenómenos da desinformação durante surtos de doenças, uma vez que pode acelerar o processo epidémico ao influenciar a resposta social, tal como é referido em [9]. Inclusive António Guterres, Secretário Geral das Nações Unidas afirma que ao mesmo tempo que estamos a lutar contra uma pandemia, também estamos a enfrentar uma epidemia de falsas informações, tal como está indicado em [10].

Desta forma foi possível identificar alguns problemas relativos à propagação de notícias falsas em época de pandemia, como o excesso de informação, por exemplo, no mês de janeiro de 2020, tal como indica Joana Gonçalves Sá, houve mais de 15 milhões de publicações no Twitter sobre a Covid-19, como é referido em [11].

Para um melhor entendimento, este trabalho está distribuido em secções da seguinte forma: na secção 2 é apresentado o estado da arte, na secção 3 está explicada a solução proposta, na secção 4 são apresentados os testes realizados, na secção 5 estão demonstrados os resultados obtidos e para finalizar as conclusões e trabalhos futuros.

## 2      Estado da Arte

Ao ser realizado um estudo sobre as notícias falsas foi possível verificar a dimensão de propagação que estas têm na sociedade atual e influência que têm nas pessoas, levando-as a ter sentimentos negativos, pois o principal dano obtido, derivado das fake news é na reputação [12]. Nesta pesquisa foram definidos e estudados alguns parâmetros para analisar o conteúdo de uma notícia com foco em países de lingua portuguesa. Após o estudo realizado foi proposto um modelo taxonómico, que foi executado com base na taxonomia desenvolvida em [13]. O modelo taxonômico proposto encontra-se dividido em quatro grupos: origem, pessoa, fonte e relação interpessoal. O parâmetro "origem", que está associado ao país de onde é proveniente a notícia, foi considerado uma problemática pois existem estudos que relacionam os países ao crescimento de propagação de fake news, sendo que há países onde estas ocorrem com mais frequência do que noutros. O exemplo disso é o inquérito realizado em [14], designado "Share of adults who have witnessed fake news in print media worldwide as of January 2019, by Country", onde é possível verificar que foram encontradas mais fake news nuns países do que noutros, ou seja, não existe um número constante definido para todos os países, sendo que existem países mais propícios à disseminação de notícias falsas do que outros.

A problemática "pessoa" refere-se ao indivíduo que partilha a notícia e encontra-se dividido em idade, educação e emprego. Relativamente à idade, foi considerado um fator problemático, na medida em que existem estudos que indicam haver idades mais prováveis para o consumo e partilha de notícias falsas, como por exemplo: "Consumming Fake news: A Matter of Age? The perception of political fake news stories in facebook ads" [15] e "Less than you think: Prevalence and predictors of fake news" [16]. A educação foi considerada uma problemática tendo por base um estudo realizado em [17], os quais consideram que quanto menor for a escolaridade de um indivíduo maior será a probabilidade de acreditar em fake news. Por último, o emprego foi considerado, devido a uma análise feita por Nicole Alvino [18], num estudo realizado empresa da qual a autora é cofundadora e diretora de estratégia; SocialChorus [1]. Nesse estudo a mesma indica que existe a necessidade de prevenir a desinformação no trabalho e ainda acrescenta que não se pode ignorar o impacto que as notícias falsas têm no local de trabalho. A fonte da notícia foi reportada, tal como se refere em [19], existem características que se devem ter em atenção ao verificar se uma notícia é falsa ou não, sendo a fonte uma delas. Neste caso, um dos pontos a ter em atenção é a credibilidade da fonte.

Relativamente às relações interpessoais, segundo as teorias do processo dual, a nossa mente inicia dois processos ao ler e receber informações, sendo uma delas superficial e automática e outra que exige esforço e concentração. Na utilização do processos superficial, o cérebro julga automaticamente a veracidade das informações com base no grau de intimidade com a pessoa que partilha a informação, ou seja, quanto maior familiaridade houver, maior será a probabilidade de uma notícia ser considerada verdadeira, mesmo que não seja, conforme descreve [20]. Tendo esses critérios, apresentamos a solução proposta conforme segue.

## 3      Solução Proposta

Desta forma, a proposta deste trabalho consiste numa ferramenta que pode auxiliar qualquer pessoa a verificar se uma notícia é falsa. Para isso foi necessário compreender o problema e desenvolver um modelo taxonômico. O modelo taxonômico desenvolvido foi dividido em quatro grupos: origem, pessoa, fonte e relação interpessoal.

O primeiro parâmetro consiste no país de origem da notícia, onde foi realizada uma pesquisa mais incidente relativa às CPLP, Comunidade de Países de Língua Portuguesa. Segundo [21], a comunidade é constituída pelos seguintes países: Angola, Brasil, Cabo Verde, Guiné-Bissau, Guiné Equatorial, Moçambique, Portugal, São Tomé e Príncipe e Timor-Leste. Estes países foram classificados como "pouco prováveis", "prováveis" e "muito prováveis" relativamente à criação ou partilha de notícias falsas, sendo que a classificação baseia-se em seis características: comparação entre o número de habitantes e o número de utilizadores da internet, comparação entre o número de utilizadores de internet e o número de utilizadores ativos nas redes sociais, existência de leis que punem a criação ou a divulgação de fake news, transparência e a confiança.

---

[1] https://socialchorus.com/team/nicole-alvino

Considerando a avaliação dada em cada uma das características, foi realizada a Tabela 1. A cada característica foi atribuída uma percentagem de 20% (100% / 5 características = 20%). Tendo em conta a avaliação determinada anteriormente, se um país for designado como muito provável numa das características será atribuído valor 0%; no caso de ser provável será atribuído valor de 10% (20% / 2 = 10%); e se for considerado pouco provável será atribuído valor de 20%, que é o valor máximo que pode ser obtido. Para concluir análise, os valores de cada linha serão somados e apresentados na última coluna, designada "Classificação final do parâmetro "origem".

Assim todos os países que apresentarem como classificação final, um valor igual ou inferior a 33.3% (100 % / 3 classificações) são considerados muito prováveis relativamente à propagação de notícias falsas. Os países que tiverem uma percentagem igual ou superior a 66.6% serão considerados países pouco prováveis. Por fim, aqueles que tiverem uma classificação final compreendida entre 33.3 e 66.6% são considerados prováveis.

**Tabela 1.** Resultado da avaliação do parâmetro "Origem"

| País | 1ª Comparação (20%) | 2ª Comparação (20%) | Leis (20%) | Transparência (20%) | Confiança (20%) | Classificação do parâmetro "origem" |
|---|---|---|---|---|---|---|
| Angola | **Provável (10%)** | **Pouco Provável (20%)** | **Pouco Provável (20%)** | **Muito Provável (0%)** | **Muito Provável (0%)** | **Provável (50%)** |
| Brasil | **Muito Provável (0%)** | **Muito Provável (0%)** | **Pouco Provável (20%)** | **Provável (10%)** | **Pouco Provável (20%)** | **Provável (50%)** |
| Cabo Verde | **Muito Provável (0%)** | **Provável (10%)** | **Pouco Provável (20%)** | **Pouco Provável (20%)** | **Muito Provável (0%)** | **Provável (50%)** |
| Guiné-Bissau | **Pouco Provável (20%)** | **Muito Provável (0%)** | **Muito Provável (0%)** | **Muito Provável (0%)** | **Provável (10%)** | **Muito Provável (30%)** |
| Guiné-Equatorial | **Pouco Provável (20%)** | **Pouco Provável (20%)** | **Pouco Provável (20%)** | **Muito Provável (0%)** | **Provável (10%)** | **Pouco Provável (70%)** |
| Moçambique | **Pouco Provável (20%)** | **Provável (10%)** | **Pouco Provável (20%)** | **Muito Provável (0%)** | **Muito Provável (0%)** | **Provável (50%)** |
| Portugal | **Muito Provável (0%)** | **Provável (10%)** | **Pouco Provável (20%)** | **Pouco Provável (20%)** | **Pouco Provável (20%)** | **Pouco Provável (70%)** |
| São Tomé e Príncipe | **Provável (10%)** | **Muito Provável (0%)** | **Pouco Provável (20%)** | **Provável (10%)** | **Muito Provável (0%)** | **Provável (40%)** |
| Timor-Leste | **Provável (10%)** | **Provável (10%)** | **Pouco Provável (20%)** | **Provável (10%)** | **Pouco Provável (20%)** | **Pouco Provável (70%)** |

O segundo grupo chama-se "pessoa" e é relativo ao indivíduo que cria ou partilha da informação, os dados recolhidos indicam nos parâmetros: idade, educação e emprego. O parâmetro idade foi "dividida" em "jovem", "idoso" e "adulto", sendo estas faixas etárias classificadas como mais prováveis ou menos prováveis, com base nas características consumo/partilha de fake news e o nº de utilizadores nas redes sociais por idade. Considerando a avaliação dada em cada uma das características, foi realizada a Tabela 2. A cada característica foi atribuído uma percentagem de 33.3% (100% / 3 características ≈ 33.3%). Tendo em conta a avaliação determinada anteriormente, se a um dos conjuntos de idade for atribuído a cor vermelha numa das características, corresponderá ao valor 0%; no caso de ser atribuído a cor laranja, corresponderá ao valor de 16.65% (33.3% / 2 = 16.65%); e se for considerado pouco provável, será atribuído a cor verde, o que corresponderá ao valor de 33.3%, que é o valor máximo que pode ser obtido. Por fim, os valores de cada linha serão somados e apresentados na última coluna, designada "Classificação final do parâmetro "Idade". Assim o conjunto que apresentar como classificação final, um valor igual ou inferior a 33.3% será considerado um grupo muito provável à propagação das fake news. O grupo que tiver uma percentagem igual ou superior a 66.6% é considerado pouco provável. Por último, aquele que tiver uma classificação final compreendida entre 33.3 e 66.6% é considerado provável.

**Tabela 2.** Classificação Final do Parâmetro Idade

| Idades | Consumo de Fake News (33.3%) | Partilha de Fake News (33.3%) | Nº de utilizadores nas redes sociais (33.3%) | Classificação parâmetro "Idade" |
|---|---|---|---|---|
| Jovem | **Verde (33.3%)** | **Verde (33.3%)** | **Vermelho (0%)** | **Verde (66.6%)** |
| Adulto | **Laranja (16.65%)** | **Laranja (16.65%)** | **Laranja (16.65%)** | **Laranja (49.95%)** |
| Idoso | **Vermelho (0%)** | **Vermelho (0%)** | **Verde (33.3%)** | **Vermelho (33.3%)** |

A educação foi dividida em três grupos: ensino básico, ensino secundário e ensino superior. De acordo com estudo realizado também foi possível identificar quem tenha o Ensino Básico é mais propício a acreditar em fake news e que quem tenha o nível de escolaridade correspondente ao ensino superior é menos propício a acreditar em notícias falsas. Desta forma a percentagem atribuída ao Ensino Básico, para fins de implementação, foi de 0% por se considerar o mais propício. Ao Ensino Superior foi de 100% por ser o nível de ensino menos propício. Por fim, ao Ensino Secundário foi atribuído o valor de 50% (100% / 2 níveis de ensino, pois estamos a excluir aquele que é definido como mínimo e tem valor atribuído de 0%). No parâmetro emprego foram consideradas as seguintes opções: desempregado, empregado no setor privado, empregado no setor público e autônomo. Emprego. Neste parâmetro as opções serão classificadas com os seguintes valores: 0%, 33.3%, 66.6% e 99.9%. Tal como já aconteceu anteriormente estes valores aproximados são obtidos através da divisão de 100% por 3, número de opções sem contar com aquele que é atribuído valor zero. Neste caso não será atribuído especificamente um valor a cada opção, pois quando forem realizados os testes, estas percentagens serão testadas em cada uma das opções, de forma a perceber de que maneira estes valores podem influenciar o resultado final obtido. O terceiro grupo abordado é a fonte da notícia está dividida em três categorias: pública (blogs, redes sociais, etc), privada (meios de comunicação) e respeitada(Governos, Militares e Polícia). Foi possível concluir que a fonte pública é mais propícia à criação/partilha de notícias falsas e que as fontes respeitadas são as menos propícias à criação/partilha de notícias falsas. Por isso às fontes públicas foi atribuído o valor de 0%, às fontes privadas foi atribuído o valor de 50% e às fontes respeitadas o valor de 100%. As percentagens anteriormente referidas foram obtidas através da divisão de 100% por dois, número de opções nas fontes, sem contar com a fonte mais propícia à partilha de notícias falsas, que lhe será atribuído valor de 0%.

As "relações interpessoais" é o último parâmetro a ser analisado e consiste na relação entre as pessoas que trocam conteúdos. Este parâmetro pode ser classificado como contato profissional, de amizade, familiar ou outro tipo de contacto. Foi possível identificar que um ambiente familiar é mais favorecedor para a partilha de notícias falsas, enquanto grupos com colegas de trabalho ou até outros grupos em que não haja uma relação de proximidade seja menos propício para a propagação de informações falsas. Neste caso foram consideradas as percentagens atribuídas no estudo que se encontra na referência [23], realizado pela Universidade de São Paulo, cuja amostra tem cerca 916 pessoas que responderam a um questionário online relativo a uma fake new que circulou na rede social WhatsApp. Dessa amostra, 51% afirma ter recebido num grupo de família, 32% num grupo de amigos, 9% em grupos de colegas de trabalho e 9% em grupos ou mensagens diretas. Deste modo o ambiente familiar é classificado com 49% (100% - 51%), amizade com 68% (100% - 32%) e as restantes classificações com 91%, sendo as percentagens atribuídas para fins de implementação. Após terem sido analisados todos os parâmetros, conseguiu-se chegar ao resultado final que se encontra na Tabela 3. Nesta tabela os parâmetros são classificados como: mínimo, médio e máximo. Se o parâmetro é classificado como mínimo significa que tem grande probabilidade de ser fake news. Se o parâmetro é classificado como médio significa que não se consegue determinar com segurança se se trata de uma fake new ou não. Se o parâmetro é classificado como máximo significa que tem grande probabilidade da notícia ser verdadeira. No caso do emprego não foi classificada nenhuma das opções pois estas irão variar durante os testes realizados, de forma a perceber de que maneira estes valores podem influenciar o resultado final obtido.

**Tabela 3.** Tabela Classificativa de parâmetros.

|  | Minimo | Médio | Máximo |
|---|---|---|---|
| País | Muito provável | Provável | Pouco provável |
| Idade | Idoso | Adulto | Jovem |
| Educação | Ensino Básico | Ensino Secundário | Ensino Superior |
| Emprego | - | - | - |
| Fonte | Pública | Privada | Respeitada |
| Relações Interpessoais | Familiar | Amizade | Contacto Profissional/ Outro tipo de contacto |

Assim ao ser selecionada uma opção por cada parâmetro será calculada uma percentagem que poderá indicar se existem uma grande probabilidade da notícia ser falsa, ou se existe uma grande probabilidade baixa da notícia ser falsa, ou se deve haver um estado de alerta, pois não é possível verificar com certezas se a notícia se trata de uma fake news ou não [22]. É importante realçar que quanto maior for a percentagem obtida, maior será a probabilidade da notícia ser verdadeira. O mesmo acontece no sentido inverso, ou seja, quanto menor for a percentagem obtida maior será a probabilidade de se tratar de uma notícia falsa. Após ser calculada a probabilidade, se a percentagem final atribuída for igual ou inferior a 44% significa que existe a probabilidade da notícia ser falsa. Se a percentagem final estiver compreendida entre 44% e 62%, significa que não existe informação suficiente para detectar se a notícia é falsa ou não, devendo o utilizador ficar atento. Por fim, caso a percentagem final seja igual ou superior a 62% significa que existe uma grande probabilidade da notícia não ser falsa. Os valores em percentagem atribuídos anteriormente resultam de um inquérito feito pela Statista [14], a 25 229 entrevistados cujo objetivo era determinar quem teria observado notícias falsas na imprensa do mundo inteiro. O resultado desta pesquisa relativa ao mundo inteiro indica que 44% encontraram notícias falsas, 39% não encontraram notícias falsas e 17% não utilizavam essas plataformas. Assim os valores inferiores ou iguais a 44% referem-se aos inquiridos que encontraram notícias falsas, os valores compreendidos entre 44% e 62% (44%+17%=61%) referem-se aos 17% de inquiridos que não utilizam essas plataformas. Os valores superiores ou iguais a 62% indicam aqueles que não encontraram notícias falsas, sendo que no final 44%+17%+39% =100%.

## 4 Testes

A análise destes dados foi realizada a partir do estudo realizado, sendo que esta trata a ferramenta de um modo geral, utilizando pesquisas realizadas, por exemplo em países diferentes dos considerados, devido ao facto de não existirem estudos realizados sobre o grupo tratado nesta ferramenta. Nestes testes foram escolhidos um país de cada uma das classificações anteriormente referidas, ou seja, Portugal que se refere a um país pouco provável, Angola que é considerado um país provável e Guiné-Bissau que é considerado um país muito provável.

Os testes realizados foram divididos em quatro fases devido também às percentagens atribuídas ao parâmetro emprego. Assim na primeira fase de testes foram atribuídas as seguintes percentagens ao parâmetro emprego: Autónomo – 0%; Desempregado – 33.3%; Privado – 66.6%; Público – 99.9%. Na segunda fase de testes foram atribuídas as seguintes percentagens: Autónomo – 99.9%; Desempregado – 0%; Privado – 33.3%; Público – 66.6%. Na terceira fase de testes foram atribuídas as seguintes percentagens: Autónomo – 66.6%; Desempregado – 99.9%; Privado – 0%; Público – 33.3%. Por fim, na quarta e última fase de testes foram atribuídas as seguintes percentagens ao parâmetro emprego: Autónomo – 33.3%; Desempregado – 66.6%; Privado – 99.9%; Público – 0%. De seguida serão apresentadas algumas análises realizadas.

Foi possível verificar, independentemente da fase de testes, que à medida que o nível educacional aumenta, a percentagem final também, indicando assim que quanto maior for o nível educacional que o indivíduo tiver maior será a probabilidade de uma notícia ser verdadeira. Também se verificou que ao manter-se os parâmetros todos iguais, variando apenas o país é possível verificar que a percentagem final é maior quanto maior for a classificação atribuída ao país, ou seja, um país pouco provável > provável > muito provável.

Dentro da empregabilidade também foi possível averiguar que ao manter-se os mesmo parâmetros selecionados, apenas variando as percentagem do emprego, é possível encontrar dois tipos de resultados: 1. Mantém-se sempre o mesmo resultado (sempre notícia verdadeira, falsa ou no estado de alerta); 2. Existe pelo menos um dos resultados diferentes, ou seja, é possível encontrar dois tipos de resultados finais.

## 5   Resultados Preliminares

Os resultados obtidos nos testes realizados na ferramenta desenvolvida, foi possível verificar e validar se uma notícia é verdadeira, falsa ou se o utilizador deve estar atento. Na Figura 1 (Fig. 1) está ilustrada a interface inicial, com alguns parâmetros selecionados. Desta forma, foi possível compreender se a percentagem final calculada é a correta com base nos valores e descrições associadas às opções selecionadas, conforme segue:

- País: Portugal (70%)
- Idade: Jovem (66,6%)
- Educação: Ensino Superior (100%)
- Emprego: Público (99,9%)
- Fonte: Respeitada (100%)
- Relação Interpessoal: Contacto Profissional (91%)

**Figura 1.** Interface Inicial

A probabilidade é calculada através da divisão da soma das percentagens associadas a cada opção selecionada pelo número total de parâmetros. Assim foi possível realizar o seguinte cálculo:

$$\frac{70\% + 66{,}6\% + 100\% + 99{,}9\% + 100\% + 91\%}{6} = \frac{527{,}5\%}{6} \approx 87{,}92\%$$

É possível concluir que a percentagem calculada anteriormente é a mesma que foi calculada pela ferramenta (Figura 2), tal como se queria provar. Na Figura 2 está ilustrada na interface de resultado final, referentes aos parâmetros que estão ilustrados na Figura 1 onde resultado final indica que se trata de uma notícia verdadeira. Nesta interface, para além de ser possível verificar qual a veracidade da notícia, também é possível voltar à interface inicial, carregando no botão "voltar". As interfaces dos restantes resultados (noticia falsa ou estado de alerta) são iguais à ilustrada, sendo que a única alteração era indicação de resultado, ou seja, indicação que se tratava de uma notícia falsa ou que o utilizador deveria estar atento.

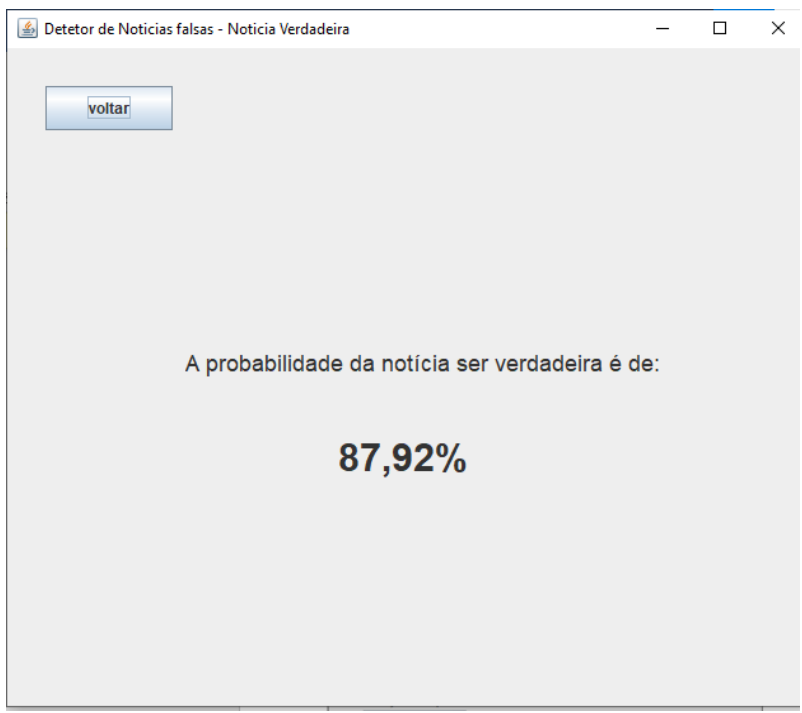

**Figura 2.** Interface de Resultado Final

## 6  Considerações Finais

Durante a realização deste trabalho a maior dificuldade verificada consistia em obter informação que relacionasse a Comunidade de Países de Língua Portuguesa (CPLP) e a temática das fake news, sendo que a existência de estudos é reduzida ou até inexistente sobre dados de alguns destes países.

A ferramenta desenvolvida é uma gota no oceano, ou seja, não influenciará o término definitivo das fake news, pois não é possível impedir que estas sejam criadas. Porém é uma grande inovação pois não existem muitas soluções associadas a esta temática, possibilitando que qualquer utilizador consiga verificar se a informação é verdadeira ou falsa, evitando que notícias falsas sejam partilhadas e consequentemente que a sociedade seja enganada. Principalmente em países onde a dúvida sobre a veracidade das informações está instalada. Consiste também numa ferramenta em que fatores como o tamanho da notícia ou até o site em que as mesmas se encontram, não seja impedimento para a utilizar, sendo uma mais valia para o utilizador estar informado.

Ao serem realizados testes, também foi possível verificar e comparar os resultados obtidos através da utilização da ferramenta, permitindo identificar se esta é um espelho da pesquisa realizada. Foram feitas comparações relativas ao país, idade e nível educacional, sendo que estas correram como o previsto. Ainda foi realizada uma comparação relativa à empregabilidade, na qual percentagens foram variadas, sendo possível identificar que existem casos onde a classificação final da notícia se mantém e outros onde a classificação da notícia pode variar.

Devido à grande quantidade de testes que teriam de ser realizados, a amostra de testes teve que ser reduzida, fazendo com que não fosse possível realizar testes para todos os parâmetros que esta ferramenta apresenta. Porém, a análise dos testes foi realizada de uma forma generalizada, pois neste caso foi possível validar a teoria pesquisada. Outra contribuição deste trabalho foi identificação e comparação entre soluções existentes, as mesmas são poucas, comparando com a grande quantidade de partilha de notícias falsas existente atualmente. Sendo esta uma temática bastante ampla, todas as oportunidades de soluções são bem-vindas. A principal contribuição deste trabalho foi a pesquisa sobre fakenews em países de lingua portuguesa e a validação com 58 resultados positivos e de acordo com os parâmtros especificados.

Em trabalhos futuros pretende-se que seja ampliada a análise de parâmetros com dados associados a cada país – Uma pesquisa, recorrendo por exemplo a questionários, de forma a tornar a ferramenta mais incidente ao impacto das fake news em cada país e não de uma forma generalizada, tal como foi implementado; A continuação da análise de testes - Realização dos testes em falta, bem como a sua análise; E que sejam Adicionados novos parâmetros – Realização de uma pesquisa, permitindo que sejam adicionadas novas classificações às já existentes, por exemplo: adicionar novos países; ou identificar novas características, de forma a aumentar o modelo taxonómico e permitir uma análise mais avançada da notícia

## Referências